\documentclass[a4paper]{article}

\usepackage{INTERSPEECH2019}
\usepackage[hyphens,spaces,obeyspaces]{url}
\usepackage{textcomp}
\usepackage{xcolor}
\usepackage{soul}
\usepackage{textgreek}
\usepackage[bottom]{footmisc}
\usepackage{graphicx}
\usepackage{amssymb,amsmath,bm}
\usepackage{textcomp}
\usepackage{graphicx}
\usepackage{wrapfig}
\usepackage{lipsum}
\usepackage{dblfloatfix}
\usepackage{fixltx2e}
\usepackage{nameref}
\usepackage{hyperref}
\usepackage{graphicx}
\usepackage{algorithm}
\usepackage{varioref}
\usepackage{booktabs} 
\usepackage{siunitx}
\usepackage[noend]{algpseudocode}
\usepackage{numprint}
\usepackage{textcomp}
\usepackage[T1]{fontenc} 
\usepackage{algorithm}
\usepackage{varioref}
\usepackage{siunitx}
\usepackage{numprint}
\usepackage{textcomp}
\def\BState{\State\hskip-\ALG@thistlm}
\def\vec#1{\ensuremath{\bm{{#1}}}}

\usepackage{lipsum}
\usepackage[noend]{algpseudocode}
\usepackage{tablefootnote}
\makeatletter
\def\BState{\State\hskip-\ALG@thistlm}
\makeatother
\usepackage{xpatch}
\xpatchcmd{\algorithmic}
{\ALG@tlm\z@}{\leftmargin\z@\ALG@tlm\z@}
{}{}
\title{Toeplitz Inverse Covariance based Robust Speaker Clustering \\
	for Naturalistic Audio Streams}
\name{Harishchandra Dubey~\thanks{\textcolor{blue}{This material is presented to ensure timely dissemination of scholarly and technical work. Copyright and all rights therein are retained by the authors or by the respective copyright holders. The original citation of this paper is: H. Dubey, A. Sangwan and J. H. L. Hansen, "Toeplitz Inverse Covariance based Robust Speaker Clustering for Naturalistic Audio Streams ", ISCA INTERSPEECH 2019, Sept. 15-19, Graz, Austria. }}, Abhijeet Sangwan, John H. L. Hansen\textsuperscript{+}~\thanks{\textsuperscript{+}This project was supported in part by AFRL under contract FA8750-15-1-0205, and partially by the University of Texas at Dallas from the Distinguished University Chair in Telecommunications Engineering held by J. H. L. Hansen.}}
\address{
Robust Speech Technologies Lab (RSTL), Center for Robust Speech Systems (CRSS)\\
The University of Texas at Dallas, 
Richardson, TX- 75080, USA
}
\email{\{Harishchandra.Dubey, Abhijeet.Sangwan, John.Hansen\}@utdallas.edu}
\begin{document}
\maketitle
%
\begin{abstract}
Speaker diarization determines~\textit{who spoke and when?} in an audio stream. In this study, we propose a model-based approach for robust speaker clustering using i-vectors. The i-vectors extracted from different segments of same speaker are correlated. We model this correlation with a Markov Random Field (MRF) network. Leveraging the advancements in MRF modeling, we used Toeplitz Inverse Covariance (TIC) matrix to represent the MRF correlation network for each speaker. This approaches captures the sequential structure of i-vectors (or equivalent speaker turns) belonging to same speaker in an audio stream. A variant of standard Expectation Maximization (EM) algorithm is adopted for deriving closed-form solution using dynamic programming (DP) and the alternating direction method of multiplier (ADMM). Our diarization system has four steps: (1) ground-truth segmentation; (2) i-vector extraction; (3) post-processing (mean subtraction, principal component analysis, and length-normalization) ; and (4) proposed speaker clustering. We employ cosine K-means and movMF speaker clustering as baseline approaches. Our evaluation data is derived from: (i) CRSS-PLTL corpus, and (ii) two meetings subset of the AMI corpus. Relative reduction in diarization error rate (DER) for CRSS-PLTL corpus is 43.22\% using the proposed advancements as compared to baseline. For AMI meetings IS1000a and IS1003b, relative DER reduction is 29.37\% and 9.21\%, respectively. 
\end{abstract}
\noindent\textbf{Index Terms}: Expectation maximization, i-Vector, Markov random field, Peer-led team learning, Speaker clustering, Toeplitz Inverse Covariance.
%
%
%
%
%
%
\vspace{-2mm}
\section{Introduction}
\label{sec:intro}
%
Speaker diarization answers "who spoke and when?" in a multi-speaker audio stream~\cite{anguera2012speaker}. Some of the practical applications of diarization technology include information retrieval~\cite{huijbregts2008segmentation}, broadcast news, meeting conversations, telephone calls, VoIP, digital audio logging~\cite{hansen2016prof} and interaction analysis in Peer-Led Team Learning (PLTL) groups~\cite{aloni2018incremental, huijbregts2012large,dubey2017csl,dubey2016slt}. Diarization is a challenging task for naturalistic audio streams as they contain short conversational turns, overlapped speech, noise and reverberation~\cite{dubey2018pltlis,hansen2018ldnn}. Variable length speech segments and unsupervised clustering step are two main challenges from modeling perspective. NIST Rich Transcription (RT) evaluations involved broadcast news data and meeting recordings for diarization study~\cite{anguera2012speaker}. Summed-channel telephone speech from NIST SRE evaluations have been used in diarization studies~\cite{aloni2018incremental}. 

Most diarization systems have following five components: (i) speech activity detection (SAD) - to remove non-speech from audio~\cite{dubey2018tasl,dubey2018icassp}; (ii) speaker change detection/segmentation; (iii) speaker modeling~\cite{dubey2019icassp,dehak2011front}; (iv) speaker clustering; and (v) re-segmentation - that re-aligns each speech frame using speaker models obtained from the clustering module.

Speaker clustering is the most important component in diarization pipeline due to which, several speaker clustering approaches have been studied. These include, but not limited to, bottom-up agglomerative hierarchical clustering (AHC)~\cite{sun2010speaker}, Cosine K-means~\cite{castaldo2008stream}, Cosine mean-shift~\cite{senoussaoui2014study}, top-down approach~\cite{meignier2006step}, integer linear programming based method~\cite{dupuy2014recent}, Viterbi-based models~\cite{lapidot2017generalized}, PLDA i-vector scoring~\cite{salmun2016use}, mixtures of von Mises-Fisher Distributions~\cite{dubey2018is}~\emph{etc}. Recent advancements in speaker clustering still remain inadequate for naturalistic audio streams such as CRSS-PLTL corpus~\cite{dubey2017csl} as those do not exploit sequential correlation in speaker turns belonging to same speaker.
In this study, we explore the correlation between different speech segments (i-vectors) belonging to same speaker for diarization task. We formulate speaker clustering as a structured inverse covariance estimation problem known as Toeplitz graphical lasso. Our work is motivated by recent success of such approaches in unsupervised analysis of physiological signals and real-world driving data~\cite{hallac2017toeplitz}. We perform comparison of proposed Toeplitz Inverse Covariance (TIC) speaker clustering with cosine K-means and movMF baseline methods~\cite{dubey2018is}. In this study, we employ i-vector features for modeling the speakers. Naturalistic audio data from CRSS-PLTL corpus is used in our experiments~\cite{dubey2016interspeech}. PLTL is a STEM education model where a peer-leader facilitate problem-solving among 6-8 students in weekly sessions~\cite{hansen2018utdallas}. We also choose two meetings subset of AMI corpus for comparative evaluations~\cite{mccowan2005ami}.  
%
%
%
\begin{figure*}[!t]
\centering
\vspace{-2mm}
\includegraphics[width=410bp]{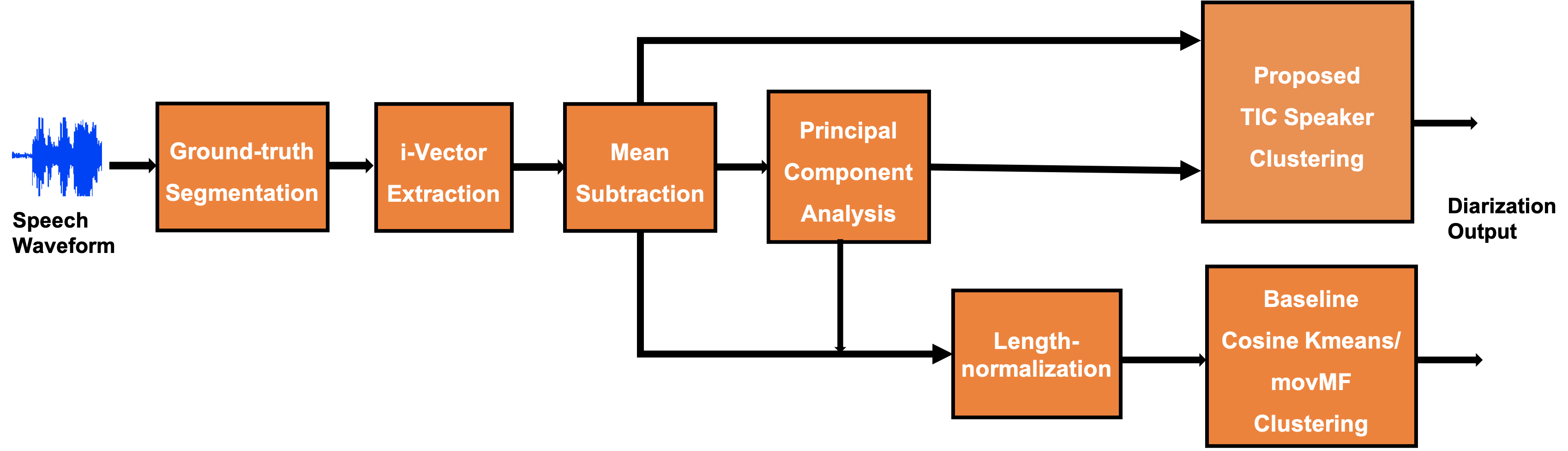}
\caption{Block diagram of diarization pipeline employing proposed Toeplitz Inverse Covariance (TIC)-based speaker clustering. We perform mean subtraction in all experiments. Length-normalization is required only for cosine K-means and movMF~\cite{dubey2018is} baselines.}
\label{fig_block_diag}
\vspace{-4mm}
\end{figure*}
\vspace{-1mm}
\vspace{-2mm}
\section{i-Vector Speaker Model}
\vspace{-2mm}
\label{sec:ivector}
Diarization involve extracting i-Vectors from short speech-segments (typically 1s) unlike speaker verification where complete utterance is used. Numerous techniques exist for clustering i-Vectors using cosine distance~\cite{senoussaoui2014study}. The i-vector framework combines the speaker and channel variability sub-spaces of linear distortion model into a total-variability space represented by matrix $\mathbf{Tmat}$~\cite{dehak2011front,hansen2015speaker}. A speaker and channel dependent Gaussian Mixture Model (GMM) supervector, $\mathbf{S}$ is decomposed as
%
\begin{equation}
\mathbf{S} = \mathbf{S_{ubm}} + \mathbf{Tmat} \cdot \mathbf{w}
\label{eqn_ivec}
\end{equation}
where $\mathbf{S_{ubm}}$ is the Universal Background Model (UBM) supervector~\cite{dehak2011front}. In Eqn.~\ref{eqn_ivec}, the latent variables, $\mathbf{w}$ are low-rank factors known as i-vectors. The total-variability matrix, $\mathbf{Tmat}$ is a low-rank projection matrix that maps high-dimensional speaker supervector to a low-dimensional total-variability space~\cite{dehak2011front,hansen2015speaker}. $\mathbf{Tmat}$ is estimated using standard expectation maximization (EM) algorithm on background speech data~\cite{kenny2005eigenvoice}. 
\setlength{\textfloatsep}{10pt}
\noindent
\begin{algorithm}[!t]
\caption{\strut \text{Assign-Clusters}}
\textbf{Input:} GIVEN $\beta \ge 0$, \textminus $LogL(i, j)$ = negative log-likelihood for i-th feature vector when it is assigned to j-th speaker cluster. $K$ is the number of speakers. Time stamp of i-vectors (speaker features) is from 1 to $T$. \\
\textbf{Output:} FinalPath i.e. cluster assignment for each i-vector. 
\begin{algorithmic}[1]	
	\Statex \hspace{-6mm} \textbf{METHOD:}		
	\State \hspace{2mm} \textbf{Initialize} \\
	\hspace{7mm} previous\_cost = list of $K$ zeros \\
	\hspace{7mm} current\_cost = list of $K$ zeros \\
	\hspace{7mm} previous\_path = list of $K$ empty lists \\
	\hspace{7mm} current\_path = list of $K$ empty lists \\
	
	\State \hspace{2mm} \textbf{for} i = 1, . . . , $T$ \textbf{do} \\
	\hspace{4mm} \textbf{for} j = 1, . . . , $K$ \textbf{do} \\
	
	\hspace{7mm} min\_index = index of minimum \{previous\_cost\} \\
	\hspace{7mm} \textbf{if} previous\_cost[min\_index] + $\beta$ > previous\_cost[j] \textbf{ then} \\
	\hspace{10mm} current\_cost[j] = previous\_cost[j] \textminus $LogL(i, j)$\\
	\hspace{10mm} current\_path[j] = previous\_path[j].append[j]
		
	\hspace{2mm} \textbf{else} \\
	\hspace{10mm} current\_cost[j] = previous\_cost[minIndex] + $\beta$ \textminus $LogL(i, j)$ \\
		
	\hspace{10mm} current\_path[j]=previous\_path[min\_index].append[j] \\
	\hspace{4mm} previous\_cost = current\_cost \\
	\hspace{4mm} previous\_path = current\_path
	\State \hspace{2mm} final\_min\_index = index of minimum \{current\_cost \} 
	\State \hspace{2mm} FinalPath = current\_path[final\_min\_index]
	\State \hspace{2mm} \textbf{return} FinalPath
	%
\end{algorithmic}
\label{assign_algo}
\end{algorithm}	
\setlength{\textfloatsep}{10pt}
\noindent
\begin{algorithm}[!t]
\caption{\strut \text{Proposed TIC-based Speaker Clustering}}	\textbf{Input:} GIVEN: (i) Algorithm~\ref{assign_algo} for assigning i-vectors to speaker clusters; (ii) i-Vectors (features) for time $1$ to $T$.
\begin{algorithmic}[1]	
\Statex \hspace{-6mm} \textbf{METHOD:}		
\State \hspace{2mm} \textbf{Initialize} \\
\hspace{7mm} Speaker cluster parameters, $\mathbf{\Theta}$ \\
\hspace{7mm} Diarization output, $spk\_labels$ = cluster assignment set $\mathbb{C}$
%
\State \hspace{2mm} \textbf{Repeat} \\
\State \hspace{7mm} \textbf{E-step:} Assign feature vectors to speaker clusters using Algorithm~\ref{assign_algo} i.e., map i-vectors $\Rightarrow$ $spk\_labels$ (see section 3.1.)\\
\State \hspace{7mm} \textbf{M-step:} Update speaker clusters (model) parameters by solving the Toeplitz Graphical Lasso (see section 3.2.) i.e., $spk\_labels$ $\Rightarrow$ $\mathbf{\Theta}$ \\
\State \hspace{2mm} \textbf{Until} Convergence
\State \hspace{2mm} \textbf{return} ($\mathbf{\Theta}$, $spk\_labels$)
\end{algorithmic}
\label{ticc_algo}
\end{algorithm}
%
\vspace{-2mm}
\section{Proposed Speaker Clustering}
\vspace{-2mm}
Toeplitz Inverse Covariance (TIC)-based clustering was found to be suitable for segmenting the real-world time-series data such as fitness-tracker and driving data~\cite{hallac2017toeplitz}. Such temporal data has complicated structure where the underlying sequences of few fixed states repeat in definitive patterns. Robust speaker clustering task possess a similar property, i.e., a small set of speakers (such as 10 for PLTL) repeat throughout the audio stream taking different conversational turns. In this section, we describe the TIC based speaker clustering followed by discussion of experimental results in next section. 

In the proposed approach, each unique speaker is represented as a correlation network modeled using MRF. Such MRF network capture the interrelationship between different i-vectors (segment-level)  belonging to corresponding (one) speaker. Thus, each speaker is modeled by their corresponding MRF. A variant of standard expectation maximization (EM) algorithm is leveraged for estimating the MRF model for each speaker. First of all, the speaker models are initialized. Next, EM iterations run alternately between Expectation (E-step) and Maximization (M-step). During the E-step, feature vectors are assigned to speaker clusters. Next in the M-step, model parameters are updated using dynamic programming (DP) and alternating direction method of multipliers (ADMM)~\cite{hallac2017toeplitz}. This process of E-step followed by M-step is repeated until convergence. Essentially, the proposed TIC approach views the temporal order of i-vectors as a sequence of states characterized by their MRF correlation network. Unlike centroid-based approaches such as cosine K-means, proposed approach models the inner structure present in the i-vector feature space. This leads to enhanced clustering performance. Each MRF models the time-invariant partial correlation structure in all feature vectors belonging to the corresponding speaker~\cite{hallac2017toeplitz}.
%
\vspace{-3mm}
\subsection{E-step: Assign feature vectors to clusters (Algorithm~\ref{assign_algo})}
\vspace{-2mm}
In this step, we fix the current value of cluster (model) parameter i.e. inverse covariance matrix, $\mathbf{\Theta}$ and solve the following combinatorial problem for obtaining the cluster assignment set, $\mathbb{C} = \{ C_{1}, C_{2}, \cdots,C_{K}\}$ given by: 
%
\begin{equation}
\text{minimize} \sum_{i=1}^{K} \sum_{X_{t} \in C_{i}} N(X_{t}, \Theta_{i}) + \beta 1_{X_{t-1} \notin C_{i}}
\label{eqn_prob5}
\end{equation}
The Eqn.~\ref{eqn_prob5} assign each of the $T$ i-vectors to one of the $K$ speaker clusters by jointly maximizing both the log likelihood and temporal smoothness. Temporal smoothness parameter $\beta$ ensures neighboring blocks are assigned to the same speaker. Thus, $\beta$ is a regularization parameter (switching penalty) that controls the trade-off between two objectives. For large value of $\beta \to \infty$, all speaker features are grouped together in a single cluster. 

Given the model parameters (i.e., inverse covariance matrices) $\Theta_{i}$'s for each of the $K$ speaker clusters, we solve the optimization problem in Eqn.~\ref{eqn_prob5} that assigns the $T$ i-vectors, $X_{1}$, $X_{2}$,....., $X_{T}$ to $K$ clusters in a way that maximizes the likelihood of data while minimizing the number of times cluster assignment changes across time. Given $K$ potential cluster assignments for each of the $T$ data points, this combinatorial optimization problem has $KT$ possible mappings of feature vectors to speaker clusters. We assign each i-vector to a unique cluster using the approach presented in the Algorithm~\ref{assign_algo}. This method is equivalent to finding the Viterbi path with minimum cost for the feature sequence of length $T$.
\vspace{-3.5mm}
\subsection{M-step: Toeplitz Graphical Lasso (Algorithm~\ref{ticc_algo})}
\vspace{-2mm}
\label{sec:lasso}
Each speaker cluster is modeled as a Gaussian inverse covariance matrix, $\Theta_{i} \in \mathbb{R}^{nbXnb}$ where $nb$ is the feature dimension. Since the inverse covariance matrix show the conditional independence structure between the variables, $\Theta_{i}$ defines a Markov Random Field (MRF) that encodes the structural representation present in all feature vectors of the i-th speaker cluster. Sparse graphical representations prevent overfitting in addition to fetching interpretable results~\cite{lauritzen1996graphical}. In the M-step of EM algorithm, our objective is to estimate the $K$ inverse covariance matrices, $\mathbf{\Theta} = \{ \Theta_{1}, \Theta_{2}, \cdots, \Theta_{K} \}$ using the cluster assignment sets, $\mathbb{C} = \{C_{1}, C_{2}, \cdots, C_{K}\}$, where $C_{i} \subset \{1, 2, . . . ,T \}$ obtained from the previous E-step. 
We can estimate each $\Theta_{i}$ in parallel thus saving execution time. The negative log likelihood of feature vector, $X_{t}$ to be assigned for i-th cluster with model parameter $\Theta_{i}$ is denoted as $N(X_{t}, \Theta_{i})$. We can write it in terms of each $\Theta_{i}$ as follows~\cite{hallac2017toeplitz}:
%
\begin{equation}
\sum_{ X_{t} \in C_{i} } N(X_{t}, \Theta_{i} ) = - |C_{i}| ( \log \text{det} (\Theta_{i}) + \text{tr}( S_{i} \Theta_{i} )) + \gamma
\label{eqn_prob41}
\end{equation}
%
%
where $|C_{i}| $ is the number of feature vectors assigned to the i-th cluster, $S_{i}$ is the empirical covariance of these feature vectors, and $\gamma$ is a constant independent of $\Theta_{i}$. Now, the M-step of EM algorithm becomes
\begin{multline}
\text{minimize} - \log \text{det} (\Theta_{i}) + \text{tr} (S_{i} \Theta_{i}) + \frac{1}{|C_{i}|} ||\mathbf{\lambda} \circ \Theta_{i} ||_{1} \\
\text{subject to  } \Theta_{i} \in \mathbb{T}
\label{eqn_prob4}
\end{multline}
%
where $\mathbb{T}$ is a set of blockwise Toeplitz matrix. Eqn.~\ref{eqn_prob4} represents a convex optimization problem known as Toeplitz graphical lasso~\cite{hallac2017toeplitz}. It is a type of graphical lasso problem with additional constraint of block Toeplitz structure on the inverse covariance matrices. Here, $\mathbb{\lambda}$ is a regularization parameter matrix for handling the trade-off between two objectives: (i) maximizing the log likelihood, and (ii) ensuring $\Theta_{i}$ to be sparse. For invertible matrix $S_{i}$, likelihood objective lead $\Theta_{i}$ to be near $S^{-1}_{i}$. The inverse covariance matrix, $\Theta_{i}$ is constrained to be block Toeplitz and $\mathbf{\lambda}$ is a $nb X nb$ matrix so that it can regularize each sub-block of $\Theta_{i}$ differently. A separate Toeplitz graphical lasso is solved for each speaker cluster at every iteration of the EM algorithm. Since we require several iterations of the EM algorithm before speaker clustering converges, a fast and efficient approach based on alternating direction method of multipliers (ADMM) is employed for this purpose~\cite{hallac2017toeplitz}. For more details on solving this Toeplitz graphical lasso, please refer to Section 4.2 in~\cite{hallac2017toeplitz}. 
%
%
%
\begin{figure}[!t]
\centering
\includegraphics[width=220bp, trim={15mm 4mm 5mm 10mm},clip]{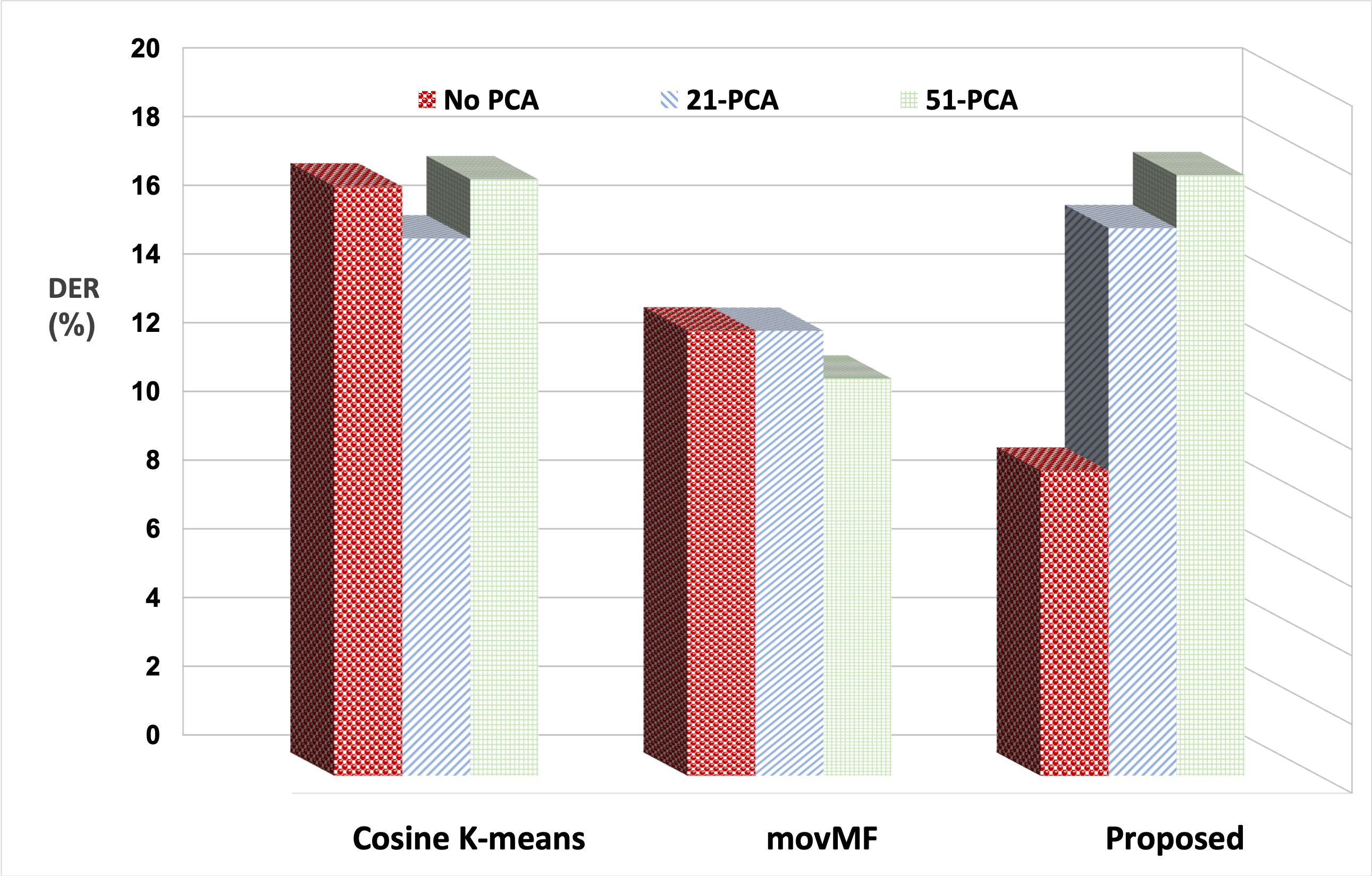}
\caption{PLTL results: DER (\%) for proposed and baseline methods. No PCA, 21-PCA and 51-PCA represent cases where no dimensional reduction is performed, where 21 principal components and 51 principal components are chosen after PCA, respectively. Proposed approach achieves significant reduction in DER as compared to both baselines.}
\label{fig_PLTL}
\end{figure}
\begin{figure}[!t]
\centering
\includegraphics[width=220bp, trim={9mm 0mm 2mm 10mm},clip]{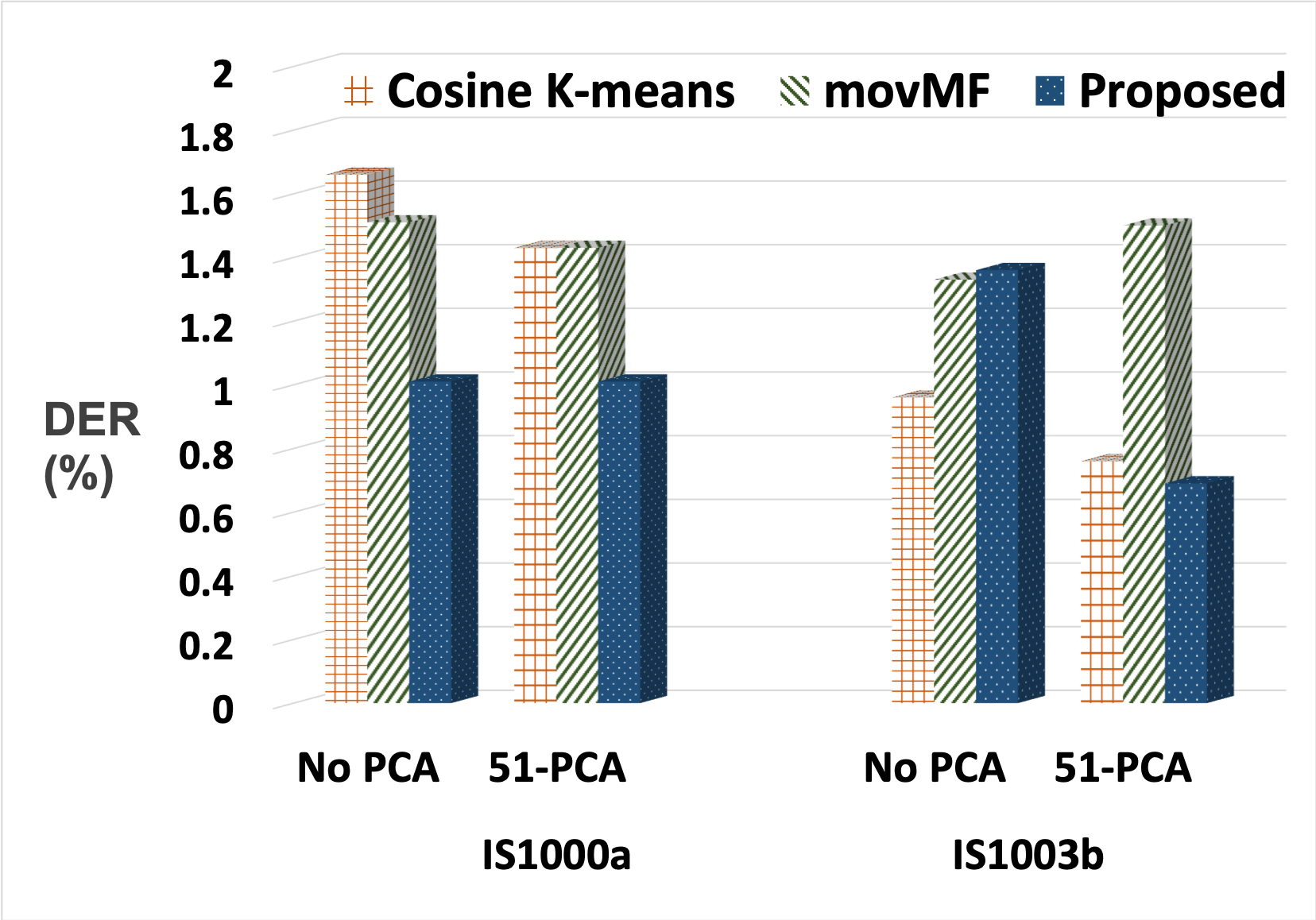}
\caption{AMI results: DER (\%) for two meetings namely IS1000a and IS1003b. No PCA and 51-PCA represent cases where no dimension reduction is performed, and 51 principal components are chosen after PCA, respectively. Proposed speaker clustering gives less than 1\% DER for both IS1000a and IS1003b (best case).}
\label{fig_AMI}
\end{figure}
\vspace{-2mm}
\section{Experiments, Results~\& Discussions}
\vspace{-2mm}
In this study, we focus on speaker clustering and hence ground-truth segmentation information is adopted for all experiments to avoid errors from SAD and speaker segmentation steps (see Fig.~\ref{fig_block_diag}). We conduct evaluations on: (i) CRSS-PLTL, and (ii) two meetings subset of AMI corpus, as detailed below. 
\vspace{-2mm}
\subsection{CRSS-PLTL Eval Set}
\vspace{-2mm}
It contain audio recordings of 80 minute  PLTL sessions that are held once per week. PLTL is a student-led STEM education paradigm popular in US universities for undergraduate courses. Each session has 6-8 students plus a peer-leader who direct the discussions for collaborative problem solving among students~\cite{carlson2016assessing}. CRSS-PLTL corpus~\cite{dubey2017csl} is a collection of audio recordings from five groups studying an undergraduate chemistry course over 11 weeks. Each participant in PLTL sessions wear a LENA device that collects the naturalistic audio~\cite{sangwan2015studying}. We choose the channel corresponding to peer-leader for single-stream diarization studies reported in this paper. Our evaluation data consists of 80 minute audio with 8 speakers. We remove the overlapped speech, laughter, cough~\emph{etc.} from audio signal before passing it through our diarization system. Majority of the conversational turns are short~(1-1.5s) for this data. Human annotators labeled the ground-truth speaker identity with start and end time stamps. 
%
%
\subsection{AMI Meeting Eval Set}
Augmented Multi-party Interaction (AMI) corpus contain audio and video data recorded from meetings. Our AMI evaluation set consists of~\textit{headset audio} of two meetings namely~\verb!IS1000a! (26 min.) and~\verb!IS1003b! (27 min.) from the popular 12-meetings subset of AMI corpus~\cite{mccowan2005ami, gonina2011fast}. Both meetings involve four speakers discussing the design of a new remote control device. 
\subsection{Evaluations~\& Baseline Systems}
Fig.~\ref{fig_block_diag} illustrate our diarization pipeline. We use frame-level 20-MFCC features extracted from 40ms windows with 10ms skip-rate. A UBM with 512 components is trained for i-vector extraction~\cite{dehak2011front}. Given the short conversational-turns in PLTL, we train an i-vector extractor of dimension 75.  For all experiments in this paper, we perform mean subtraction on i-vectors where the mean is computed from entire PLTL session/AMI meeting (see Fig.~\ref{fig_block_diag}). Mean subtraction is followed by principal component analysis (PCA) for de-correlation and dimension reduction. We do PCA based dimension reduction to compare the performance over 21 and 51 components. 

We recently proposed Mixtures of von Mises-Fisher Distributions (movMF) based speaker clustering for naturalistic audio data~\cite{dubey2018is}. In this study, we choose cosine K-means~\cite{zhong2005efficient} and movMF speaker clustering~\cite{dubey2018is} as our baseline approaches. Unlike conventional K-means, cosine K-means projects the estimated cluster centroids onto unit hypersphere at the end of each M-step in the EM algorithm. Cosine K-means is a special case of movMF speaker clustering as discussed in~\cite{dubey2018is}. When all mixture weights are equal and all concentration parameters are equal, the movMF speaker clustering becomes equivalent to cosine K-means. Feature vectors undergo length-normalization~\cite{garcia2011analysis} before passing through cosine K-means~\cite{zhong2005efficient} and movMF~\cite{dubey2018is} baselines as these methods expect data lying on the unit hypersphere. On the other hand,  length-normalization is optional for proposed TIC speaker clustering. We input either the mean-subtracted i-vectors or PCA processed i-vectors to the proposed TIC speaker clustering. 
%
%
\subsection{Results \& Discussions}
%
Diarization error rate (DER) is used for comparing the performance of different clustering approaches in this paper~\cite{nistder}. Mathematically, it is defined as:
%
\begin{equation}
\text{DER} = \frac{\alpha_{fa} + \alpha_{miss} + \alpha_{err} }{\alpha_{total}}
\label{eqn_der}
\end{equation}
where $\alpha_{total}$ is the total scored time from reference annotations, $\alpha_{fa}$ is scored time for which a non-speech region is incorrectly marked as containing speech, $\alpha_{miss}$ is scored time for which a speech region  is incorrectly marked as not containing speech, and $\alpha_{err}$ is the scored time for which a wrong  speaker identity is assigned within a speech region. In this study, we focus on speaker clustering and hence ground-truth segmentation information was adopted in all experiments. Thus, $\alpha_{fa}$ and $\alpha_{miss}$ are zero. Here, DER is only dependent on speaker confusion error, $\alpha_{err} $. Overlapped speech, laughter and other miscellaneous vocalizations such as cough is removed from the audio signal before feeding into our diarization pipeline (see Fig.~\ref{fig_block_diag}).

%
%
The proposed speaker clustering approach requires $\mathcal{O}(KT)$ time for grouping $T$ i-vectors into $K$ clusters. Hence, it is a scalable approach.  Fig.~\ref{fig_PLTL} shows the DER comparison between proposed and baseline methods with various post-processing for PLTL data. Clearly, DER is reduced to almost half for proposed approach as compared to cosine K-means. The proposed method learns a correlation network modeled as MRF and hence it expects correlated data as input. After PCA, feature vectors are de-correlated and estimating the underlying correlation model using MRF is not accurate in this case. Consequently, as expected PCA degrades the performance of proposed approach on naturalistic PLTL data. The movMF is state-of-the-art for speaker clustering on PLTL data~\cite{dubey2018is}. The proposed approach lead to 23.18\% relative reduction in DER as compared to movMF speaker clustering. Significant DER reduction validate the efficacy of proposed Toeplitz Inverse Covariance (TIC)-based speaker clustering for naturalistic audio such as CRSS-PLTL.

Fig.~\ref{fig_AMI} illustrate the DER comparison between proposed approach and baselines for two meetings subset of AMI corpus. Clearly, the best DER for both meetings are obtained by proposed approach. For both meetings, proposed approach fetch less than 1\% DER showing its efficacy for real-world meeting data.  For AMI meetings IS1000a and IS1003b, relative DER reduction is 29.37\% and 9.21\%, respectively. CRSS-PLTL data has higher levels and more varied forms of distortions as compared to AMI meeting data. Also, there are 8 speakers in PLTL data as compared to 4 in AMI meetings. These differences cause the DER to be in different ball park for both corpora.
\vspace{-2mm}
\section{Summary~\& Conclusions}
\vspace{-1mm}
In this paper, we leveraged the Toeplitz Inverse Covariance (TIC) estimation in speaker clustering task for naturalistic audio such as CRSS-PLTL corpus. Such audio data are rich in noise, overlapped-speech and reverberation in addition to short conversational turns (1s). The proposed approach accurately models the inner structure in i-vector features belonging to same speaker. It leads to enhanced clustering that eventually improves diarization performance. Each speaker cluster is characterized by a Markov Random Field (MRF) that captures its correlation structure. For robust speaker clustering task, the proposed approach iterates with alternating two steps of the standard Expectation Maximization (EM) algorithm: (i)~\textbf{E-step} that assigns each speaker feature to a unique cluster; and (ii)~\textbf{M-step} that update the model parameters using speaker clusters obtained in the E-step. We performed experiments on (i) CRSS-PLTL; and (iii) AMI meeting data. Cosine K-means and movMF were chosen as baseline clustering methods. Proposed TIC based speaker clustering fetch 43.22\% relative reduction in DER for CRSS-PLTL data. AMI evaluation set has 29.37\% (IS1000a) and 9.21\% (IS1003b) relative reduction in DER over baseline. Significant improvements in DER show the efficacy of proposed speaker clustering for naturalistic audio streams.
\bibliographystyle{IEEEtran}
\bibliography{INTERSPEECH2019}
\end{document}